\newcommand{\bbweathd}{\emph{bbweathd}}
\newcommand{\bbd}{\emph{bbd}}
\newcommand{\braind}{\emph{braind}}
\newcommand{\spotd}{\emph{spotd}}
\newcommand{\weathd}{\emph{weathd}}

\newcommand{\torusd}{\emph{torusd}}
\newcommand{\guided}{\emph{guided}}
\newcommand{\clamd}{\emph{clamd}}
\newcommand{\focusd}{\emph{focusd}}
\newcommand{\taosd}{\emph{taosd}}
\newcommand{\teld}{\emph{teld}}
\newcommand{\grbd}{\emph{grbd}}
\newcommand{\forwarder}{\emph{forwarder}}
\newcommand{\tlat}{\theta_\mathrm{lat}}
\newcommand{\hobs}{h_\mathrm{obs}}
\newcommand{\dobs}{\delta_\mathrm{obs}}

\newcommand{\fig}[1]{\mbox{Figure\hspace{0.2em}\ref{#1}}}

\newcommand{\sect}[1]{\mbox{\S\ref{#1}}}
\newcommand{\tbl}[1]{\mbox{Table\hspace{0.3em}\ref{#1}}}

\documentclass[12pt,preprint]{aastex}

\begin{document}

\title{The Taiwanese-American Occultation Survey: The Multi-Telescope Robotic
Observatory}
\author{M.~J.~Lehner\altaffilmark{1,2,3},
C.-Y.~Wen\altaffilmark{1},
J.-H.~Wang\altaffilmark{1,7},
S.~L.~Marshall\altaffilmark{4,5},
M.~E.~Schwamb\altaffilmark{6},
Z.-W.~Zhang\altaffilmark{7},
F.~B.~Bianco\altaffilmark{2,3},
J.~Giammarco\altaffilmark{8},
R.~Porrata\altaffilmark{9},
C.~Alcock\altaffilmark{3},
T.~Axelrod\altaffilmark{10},
Y.-I.~Byun\altaffilmark{11},
W.~P.~Chen\altaffilmark{7},
K.~H.~Cook\altaffilmark{5},
R.~Dave\altaffilmark{12},
S.-K.~King\altaffilmark{1},
T.~Lee\altaffilmark{1},
H.-C.~Lin\altaffilmark{7},
S.-Y.~Wang\altaffilmark{1},
J.~A.~Rice\altaffilmark{13} and
I.~de~Pater\altaffilmark{14}
}
\altaffiltext{1}{Institute of Astronomy and Astrophysics, Academia Sinica.
 P.O. Box 23-141, Taipei 106, Taiwan}\email{mlehner@asiaa.sinica.edu.tw}
\altaffiltext{2}{Department of Physics and Astronomy, University of
 Pennsylvania, 209 South 33rd Street, Philadelphia, PA 19104}
\altaffiltext{3}{Harvard-Smithsonian Center for Astrophysics, 60 Garden Street,
 Cambridge, MA 02138}
\altaffiltext{4}{Kavli Institute for Particle Astrophysics and Cosmology,
 2575 Sand Hill Road, MS 29, Menlo Park, CA 94025}
\altaffiltext{5}{Institute for Geophysics and Planetary Physics, Lawrence
 Livermore National Laboratory, Livermore, CA 94550}
\altaffiltext{6}{Division of Geological and Planetary Sciences,
 California Institute of Technology, 1201 E. California Blvd.,
 Pasadena, CA 91125}
\altaffiltext{7}{Institute of Astronomy, National Central University, No. 300,
 Jhongda Rd, Jhongli City, Taoyuan County 320, Taiwan}
\altaffiltext{8}{Department of Astronomy and Physics,
 Eastern University 1300 Eagle Road Saint Davids, PA 19087}
\altaffiltext{9}{Department of Physics, University of California at Berkeley, 
 366 Le Conte, Berkeley, CA 94270}
\altaffiltext{10}{Steward Observatory, 933 North Cherry Avenue, Room N204
 Tucson AZ 85721}
\altaffiltext{11}{Department of Astronomy, Yonsei University, 134 Shinchon,
 Seoul 120-749, Korea}
\altaffiltext{12}{Initiative in Innovative Computing, Harvard University,
 60 Oxford St, Cambridge, MA 02138}
\altaffiltext{13}{Department of Statistics, University of California Berkeley,
367 Evans Hall, Berkeley, CA 94720}
\altaffiltext{14}{Department of Astronomy, University of California Berkeley,
 601 Campbell Hall, Berkeley CA 94720}

\begin{abstract}
The Taiwanese-American Occultation Survey (TAOS) operates four fully
automatic telescopes to search for occultations of stars by Kuiper
Belt Objects. It is a versatile facility that is also useful for the
study of initial optical GRB afterglows. This paper provides a
detailed description of the TAOS multi-telescope system, control
software, and high-speed imaging.
\end{abstract}

\keywords{Solar System, Astronomical Instrumentation, Astronomical Techniques }

\section{Introduction}
\label{sec:intro}

The Taiwanese American Occultation Survey (TAOS) continuously monitors
up to $\sim$1,000 stars with four telescopes at $5$~Hz, for the
purpose of detecting occultations of these stars by small
($\gtrsim$1~km) objects in the Kuiper Belt and beyond. Installation of
the first three telescopes and the first version of the control
software were completed in January 2005, and the survey was officially
started shortly thereafter. Installation of the fourth telescope was
completed in January 2007, and four telescope operations commenced in
August 2008. The TAOS Collaboration has finished analysis of the first
two years of three-telescope data, which comprises over
150,000~star-hours of observations. The results of this analysis are
presented in \citep{kiwiapjl}.

The principal science questions addressed with this survey are:
\begin{itemize}
\item What are the number and size distribution of small ($\sim$0.5--10~km in
 diameter) bodies in the trans-Neptunian region?
\item Is there an extension of the Kuiper Belt beyond $50$~AU
  comprising bodies too small to have been detected in direct surveys?
\item What are the number and size distribution of objects (e.g. Sedna)
  in the \emph{Extended Disk} of the Solar System?
\end{itemize}

The signatures we are searching for are brief reductions in flux from
background stars due to occultations by small Kuiper Belt Objects
(KBOs) and Extended Disk Objects (EDOs). The occultation survey
technique is uniquely suited to the detection of such objects, since it
is capable of detecting objects that are much fainter than the
magnitude limit of any plausible direct survey
\citep{1976Natur.259..290B}. However, the occultation technique is
difficult to implement in practice, so the project includes some
important technology development goals. For more details on the
occultation technique applied to searches for outer Solar System
objects see \citet{2008AJ....135.1039B}, \citet{2007AJ....134.1596N},
\citet{cooray}, \citet{2003ApJ...587L.125C},
\citet{2006AJ....132..819R}, \citet{2007MNRAS.378.1287C}, and
\citet{2003ApJ...594L..63R}. For an overview of the TAOS Project, see
\citet{2006AN....327..814L}, \citet{2007IAUS..236...65C}, and
\citet{2003EM&P...92..459A}.  For more information on the Kuiper Belt,
see \citet{1993Natur.362..730J}, \citet{2002ARA&A..40...63L},
\citet{2004AJ....128.1364B}, and \citet{2001AJ....122.2740T}.  For
more information on the Extended Disk, see
\citet{2004ApJ...617..645B}, \citet{2004Natur.432..598K},
\citet{2004AJ....128.2564M}, and \citet{2007Icar..191..413B}.

In addition to the primary science goals stated above, a rapid
response capability was implemented on the TAOS system to observe
optical counterparts to Gamma Ray Burst (GRB) events reported to the
GRB Coordinates Network\footnote{\url{http://gcn.gsfc.nasa.gov/}, see
  also \citet{1998AIPC..428...99B}} (GCN). The TAOS system observed
its first optical counterpart to a GRB in October of 2007
\citep{grb}. The follow-up capability will be described in
\sect{sec:grb}.

\subsection{Design Criteria}
\label{sec:design}

Implementation of an occultation survey for detection of small KBOs
presents a number of challenges. First, such events have very short
durations, typically about 200~ms \citep{2007AJ....134.1596N}. Second,
such events are extremely rare. Estimates of occultation event rates
are model dependent and range from as low as $10^{-4}$~events per star
per year \citep{2004AJ....128.1364B} to as high as $10^{-2}$~events
per star per year \citep{panandsari}. (Note that when this survey was
designed in 1996, occultation event rates as high as $10^{-1}$~per
star per year were predicted for objects down to 1~km in diameter, but
the results of \citet{2004AJ....128.1364B} showed a likely break in
the size spectrum at diameters of about 30~km, which lowered the
expected event rate.)

The TAOS system was thus designed to meet the following criteria:
\begin{itemize}
\item Make photometric measurements at a high sampling rate (5~Hz).
\item Continuously monitor enough stars ($\sim$1,000) to obtain a
  significant event rate.
\item Low ($\lesssim$0.1~ per year) false positive rate.
\end{itemize}
The rapid photometry is achieved by means of the innovative use of an
otherwise conventional CCD camera (which will be described in
\sect{sec:zipper}). An adequate number of stars can be imaged by a
modern CCD camera mounted on a small (50 cm), fast (F/1.9) telescope
with a wide field of view (3$\Box^\circ$).

Achieving the goal of a low false positive rate was a primary driver
of the design. Our goal was to be able to measure a rate of stellar
occultations by KBOs of 1~event per 1,000~stars per year of observing
time.  We make $\sim$$10^{10}$ photometric measurements per year, from
which we want to derive less than one spurious occultation.  The false
alarm probability per observation must therefore be exceedingly small
if the results of the survey are to be interpreted with confidence.
We control false alarms by requiring joint detections in an array of
four independent robotic telescopes. False positives of terrestrial or
atmospheric origin (such as birds, aircraft, extreme atmospheric
scintillation events, etc.)  are removed by this requirement.  False
positives due to main belt asteroids can be eliminated by automated
follow-up of detected events at a nearby, larger telescope. However,
this would require a real-time analysis pipeline, which has not yet
been implemented for this project. The telescopes are spaced far
enough apart that scintillation events are unlikely to affect all of
the telecopes simultaneously. Note, however, that the telescopes are
close enough that any occultation events would be detected by all four
telescopes (see \citet{2007AJ....134.1596N} for a detailed description
of occultation events from objects in the Outer Solar System).
Finally, note that the redundancy in the multi-telescope system also
greatly enhances the credibility of the reported results. It would be
difficult to credibly report on rare occultation events that were seen
in only a single telescope.

This paper will specifically address the first two design criteria
listed above. A detailed description of the survey system (both
hardware and software) will be presented, and the implementation of
the survey system to address the design criteria will be discussed. We
have developed statistical analysis techniques which make efficient
use of the multi-telescope data to limit the false positive rate and
provide a credible result, however, discussion of these techniques is
outside the scope of this paper. An introduction to the statistical
analysis methods used by the TAOS project can be found in
\citet{chyng}, and a discussion of the application of these techniques
will appear in a future paper \citep{stat}. 

\section{The Site}
\label{sec:site}
The telescopes are installed atop Lu-Lin Mountain (longitude
$120\degr~50\arcmin~28\arcsec$~E; latitude $23\degr~30\arcmin$~N,
elevation 2850~meters), in the Yu Shan (Jade Mountain) area of Taiwan.
The median seeing is $1.3\arcsec$, with $\sim$100~clear nights per
year.  Most of the poor nights occur during rainy season (May through
September). Since the beginning of 2005, TAOS has averaged $\sim$400 hours
per year of actual observing, an average of four hours per clear
night. High humidity is typically the reason the observing hours are
limited on clear nights. (Plans are currently being developed to start
a new survey at a drier site.)

\begin{figure}[bt]
\plotone{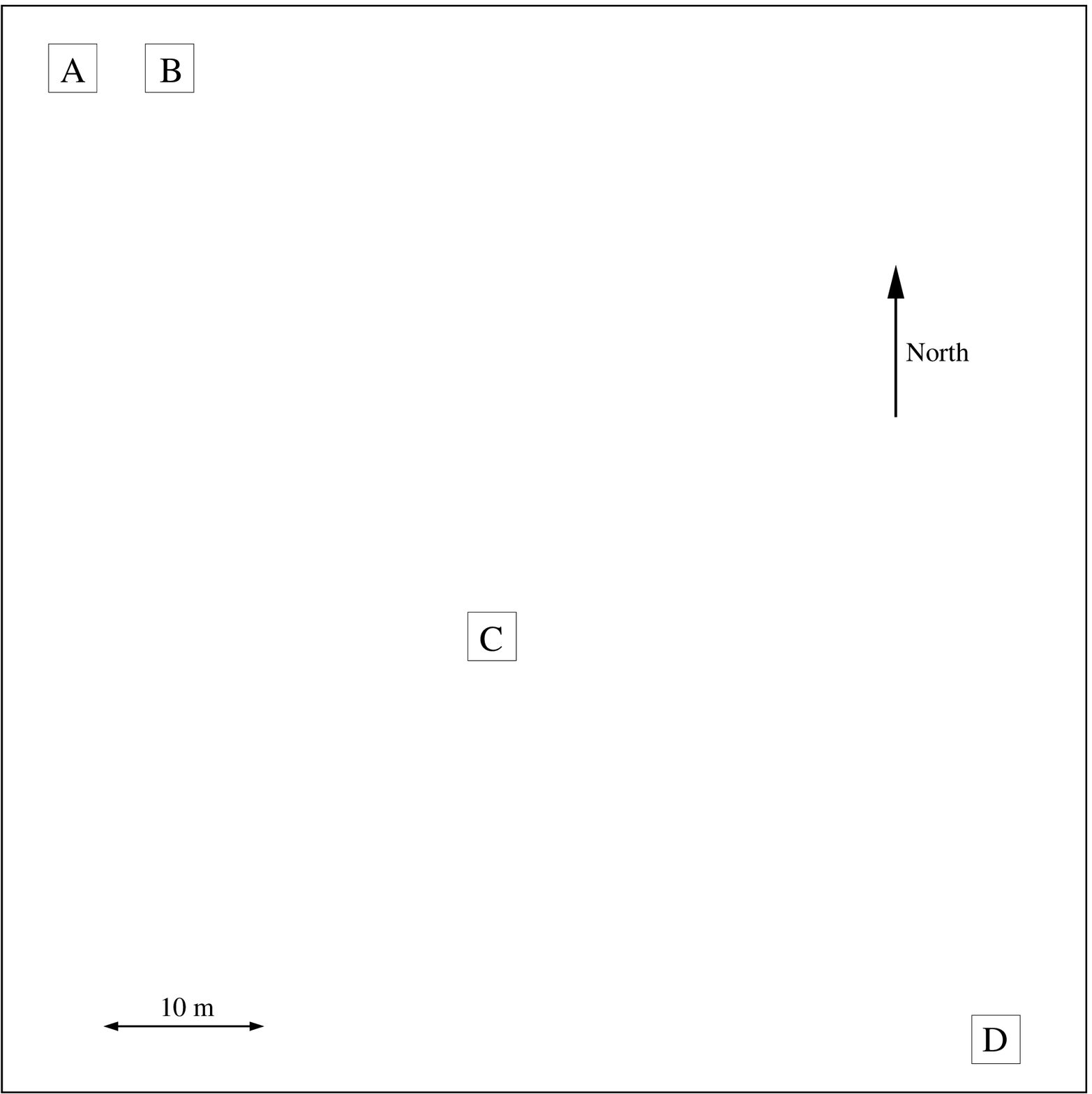}
\caption[]{Layout of the four TAOS telescopes at Lu-Lin Observatory.}
\label{fig:site}
\end{figure}

The telescopes have separations ranging from 6~meters to 60~meters
(see \fig{fig:site}). These separations are sufficient to eliminate
most, if not all, false positives arising from extreme scintillation
events. Note that the original design for the TAOS project had the
fourth telescope sited at a different peak located 7~km to the west of
the Lu-Lin site \citep{2001ASPC..246..253K}. The reason for this was
that differences in the timing of an event would give a measurement of
the relative velocity of the occulting object, thus allowing the
object distance to be estimated. This plan was abandoned for several
reasons. First, the development of the second site was logistically
very difficult, and hence extremely expensive. Second, an occultation
shadow crossing the Lu-Lin site would cross the second site only
during part of the year, due to the obliquity of the Earth. Third, our
5~Hz sampling rate is not nearly fast enough to give a velocity
measurement sufficiently accurate to provide a reasonable estimate of
the distance. Finally, it was determined that a four-telescope system
would provide significantly better statistical identification of false
positives than a three-telescope system, as well as enable detection
of smaller objects. This will be described in \citet{stat}.

\section{The Hardware}

\subsection{The Telescopes}
\label{sec:telescopes}

\begin{figure}[bt]
\plotone{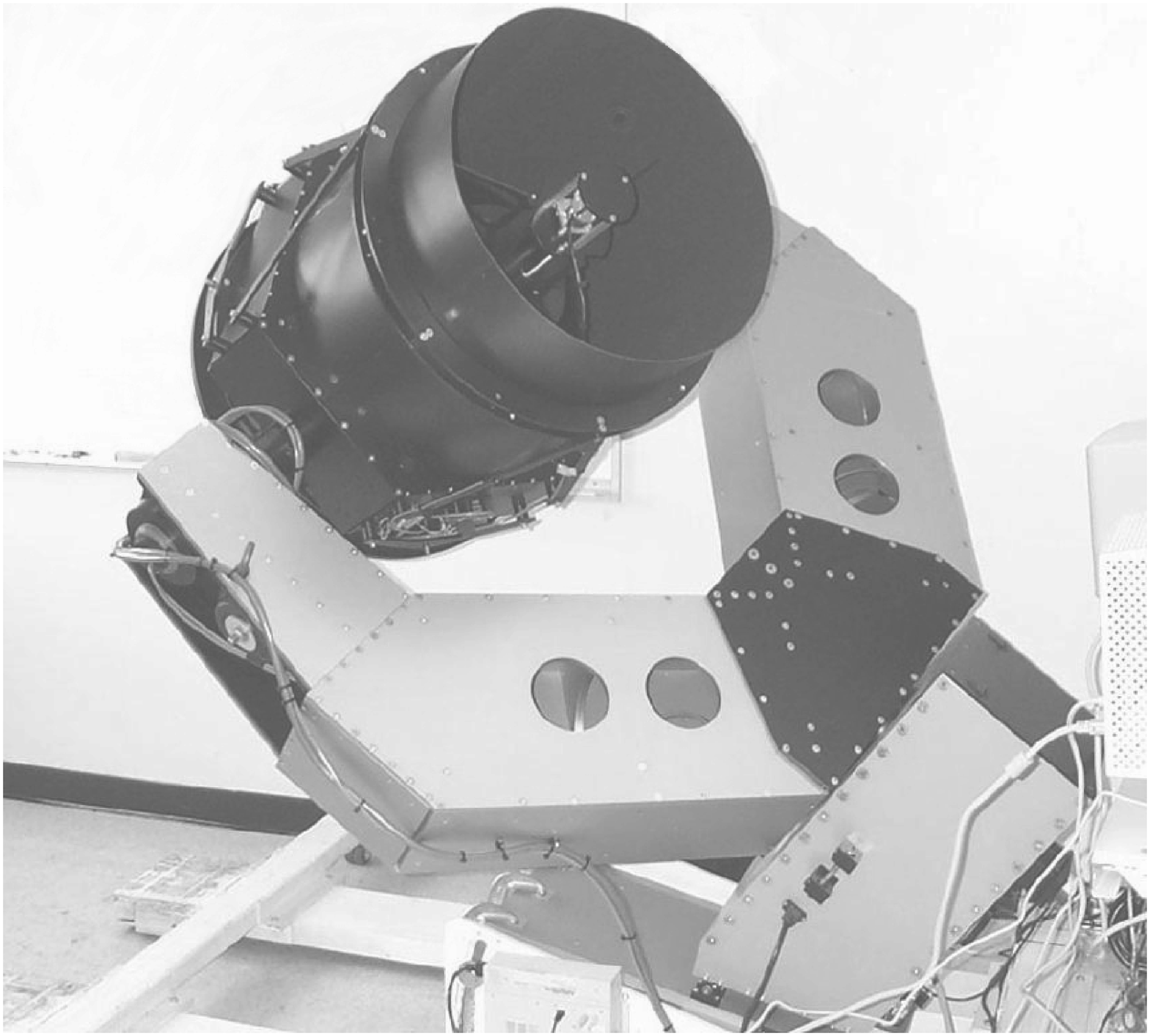}
\caption[]{The TAOS~B telescope.}
\label{fig:telescope}
\end{figure}

The four identical telescopes were manufactured by Torus Technologies
(now Optical Mechanics, Inc.). The TAOS telescopes were designed to be
compact, and the optical design was based on the system requirement to
cover a large number of stars using a camera with a single CCD
chip. Considering the affordable cost and ease of manufacturing, a
50~cm aperture was chosen.  Each telescope is a Cassegrain system with
the focal plane located behind the primary mirror. In order to provide
the fast converging beam for a wide field of view, a parabolic primary
mirror with F/1.5 is used. Combined with a large F/3.6 spherical
secondary mirror (25~cm diameter), the overall F~ratio is about
1.9~with an effective field of view of 3$\Box^\circ$. With a 14~$\mu$m
pixel CCD, the plate scale is about 3~arc~sec/pixel, which
undersamples the seeing of the site. The effective collecting area is
about 1875~cm$^2$.The wide field correctors consist of five lenses to
provide a flat focal plane. Four different materials were used for the
chromatic correction. The design criterion was to provide 60\% of the
enclosed energy within one pixel area within the passband between
500~to 750~nm. The design and modeling of the telescope was done using
the ZEMAX\texttrademark software
package\footnote{\url{http://www.zemax.com/}}. The optical layout is
shown in \fig{fig:telescope} and the detailed parameters for the
design are summarized in \tbl{tab:optics}.

\begin{figure*}[bt]
\plotone{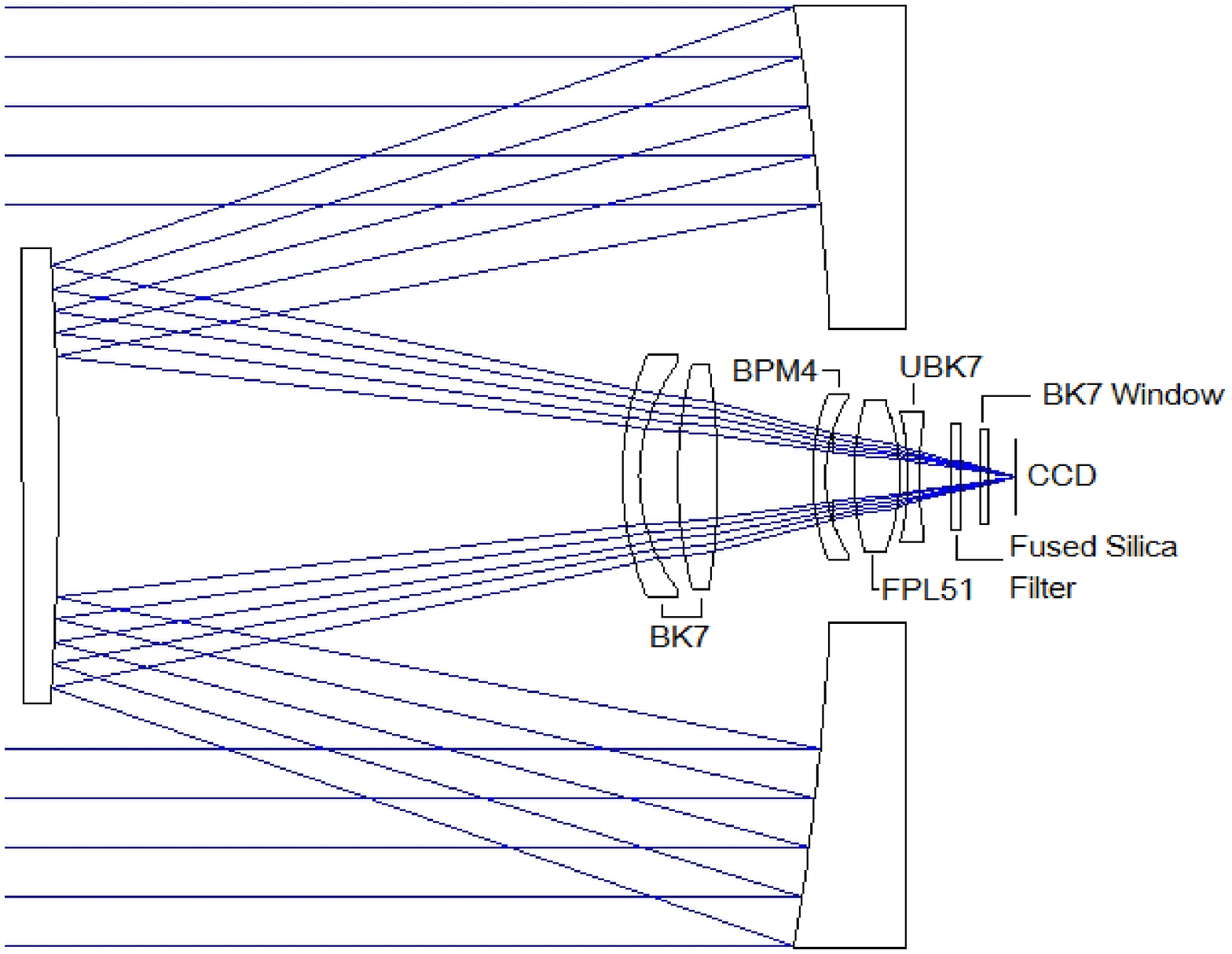}
\caption[]{Optical layout of the TAOS telescopes.}
\label{fig:optics}
\end{figure*}

\begin{deluxetable}{lrrrrl}
\rotate
\tabletypesize{\small}
\tablecolumns{6}
\tablewidth{0pc}
\tablecaption{Optical components of the TAOS telescopes}
\tablehead{Element & Radius of Curvature~(mm) & Axial Thickness~(mm) &
 Clear Aperture~(mm) & Conic Constant & Material}
\startdata
Primary mirror   & -1496.332 &          & 505.910 & -1 & Pyrex  \\
                 &           & -418.142 &         &    & Air    \\
Secondary mirror & -1795.977 &          & 244.953 &  0 & Pyrex  \\
                 &           &  304.360 &         &    & Air    \\
Lens 1           &   165.470 &    9.000 & 130.467 &  0 & BK7    \\
                 &    98.600 &   20.778 & 121.809 &  0 & Air    \\
Lens 2           &   239.650 &   21.356 & 121.570 &  0 & BK7    \\
                 &  -408.100 &   51.233 & 119.782 &  0 & Air    \\
Lens 3           &   124.500 &    7.000 &  89.810 &  0 & BPM4   \\
                 &    77.790 &   15.419 &  83.550 &  0 & Air    \\
Lens 4           &   142.550 &   24.535 &  81.705 &  0 & FPL51  \\
                 &  -108.870 &    4.522 &  77.973 &  0 & Air    \\
Lens 5           &  -149.580 &    6.000 &  69.911 &  0 & UBK7   \\
                 &   198.530 &   17.030 &  64.277 &  0 & Air    \\
Filter           &  $\infty$ &    5.000 &  57.061 &  0 & BK7    \\
                 &  $\infty$ &   10.775 &  55.433 &  0 & Air    \\
Camera window    &  $\infty$ &    4.000 &  50.037 &  0 & Silica \\
                 &  $\infty$ &   15.329 &  48.681 &  0 & Vacuum \\
CCD detector     &  $\infty$ &          &  41.070 &  0 &        \\
\enddata
\label{tab:optics}
\end{deluxetable}

Each telescope is controlled by an Oregon Micro Systems PC68 card. The
HA and Dec axes are driven by friction wheels which are powered by
Bearing Engineers Inc. Advanced Vector Servo systems. The axis
positions are monitored by Renishaw RGH24 rotary encoders with
0.27~arc~sec/step resolution. The secondary focus position is set by a
screw drive powered by an Oriental Vexta PK245-03B motor controlled by
an Intelligent Motion Systems IB462 bipolar stepping motor driver.

All of the telescopes were delivered with strong spherical aberrations
over the whole field as well as significant astigmatism in the corners
of the images. The primary mirror on the fourth telescope was found to
have very severe surface defects which distorted the point spread
function beyond any reasonable shape and size.  Each of the telescopes
was also found to have severe mechanical and defects. We found the
mirrors would move by large amounts (the primary mirrors would move
\hbox{4--5}~mm as the telescope would slew from horizon to horizon,
while the specified tolerance was $<50$~$\mu$m). We also found the
tracking accuracy limited by backlash in the drive system reduction
gears and bad bearings in the friction wheel assemblies. The secondary
focus systems were found to have poor resolution in positioning and
significant backlash.

We have made significant improvements to the telescopes such that they
are now of sufficient quality for the requirements of the survey.
First, in order to keep the mirrors accurately positioned, the entire
mirror support structures for both the primary and secondary mirrors
were redesigned and refabricated by engineers at Lawrence Livermore
National Laboratory.  Second, the friction wheel assemblies were
redesigned and refabricated by engineers at the Institute for
Astronomy and Astrophysics at Academia Sinica in Taiwan, and the
reduction gears were replaced with Bayside model PX-34-050-LB (50:1)
devices.  Third, the secondary focus drive systems were completely
redesigned by engineers at Academia Sinica. A Renishaw RGH22 linear
encoder with 0.5~$\mu$m resolution was added to each telescope to
enable the secondary focus positions to be accurately set to within
1.25~$\mu$m (the limiting resolution from the stepper motor and
threading of the drive shaft). The encoders are equipped with limit
switches to keep the secondary mirror from bottoming out against the
support structure in one direction and completely unscrewing from the
support structure in the other direction. The encoder system is also
equipped with a magnetic reference mark to reproduce the home position
within one motor step, which is needed for the automatic focus control
(see \sect{sec:focus}).  Finally, the primary mirror of the fourth
telescope was repolished and resurfaced by the Space Optics Group at
the Korea Research Institute of Standards and Science in Daejeon,
South Korea, and it is now of quality comparable to the three other
primaries.

Due to the poor optical quality of the optical components, the actual
enclosed energy within one pixel averages about 30\% rather than the
design goal of 60\%.  The best PSFs now range from 1.6~pixels (4.6~arc
sec) in the center of the image to 2.5~pixels (7.3~arc sec) in the
corners.  This lowered the limiting magnitude of the survey. We have
partially compensated for the reduced optical quality by an upgrade to
the CCD camera from a low quantum efficiency, front illuminated device
to a camera with a back illuminated CCD (as will be described in the
following section). The limiting magnitude of the survey is now
roughly $R=13.5$, rather than our target of $R=15$. We have made a
careful search for the densest fields along the ecplitic (see
\sect{sec:fields}), and we are typically able to monitor fields with
300 -- 500 stars, which is within a factor of 2 -- 3 of our design
goal. We have designed a Hartmann-Shack device to routinely check on
the optical performance of telescopes.

\subsection{The Cameras}
\label{sec:camera}

\begin{figure}[bt]
\plotone{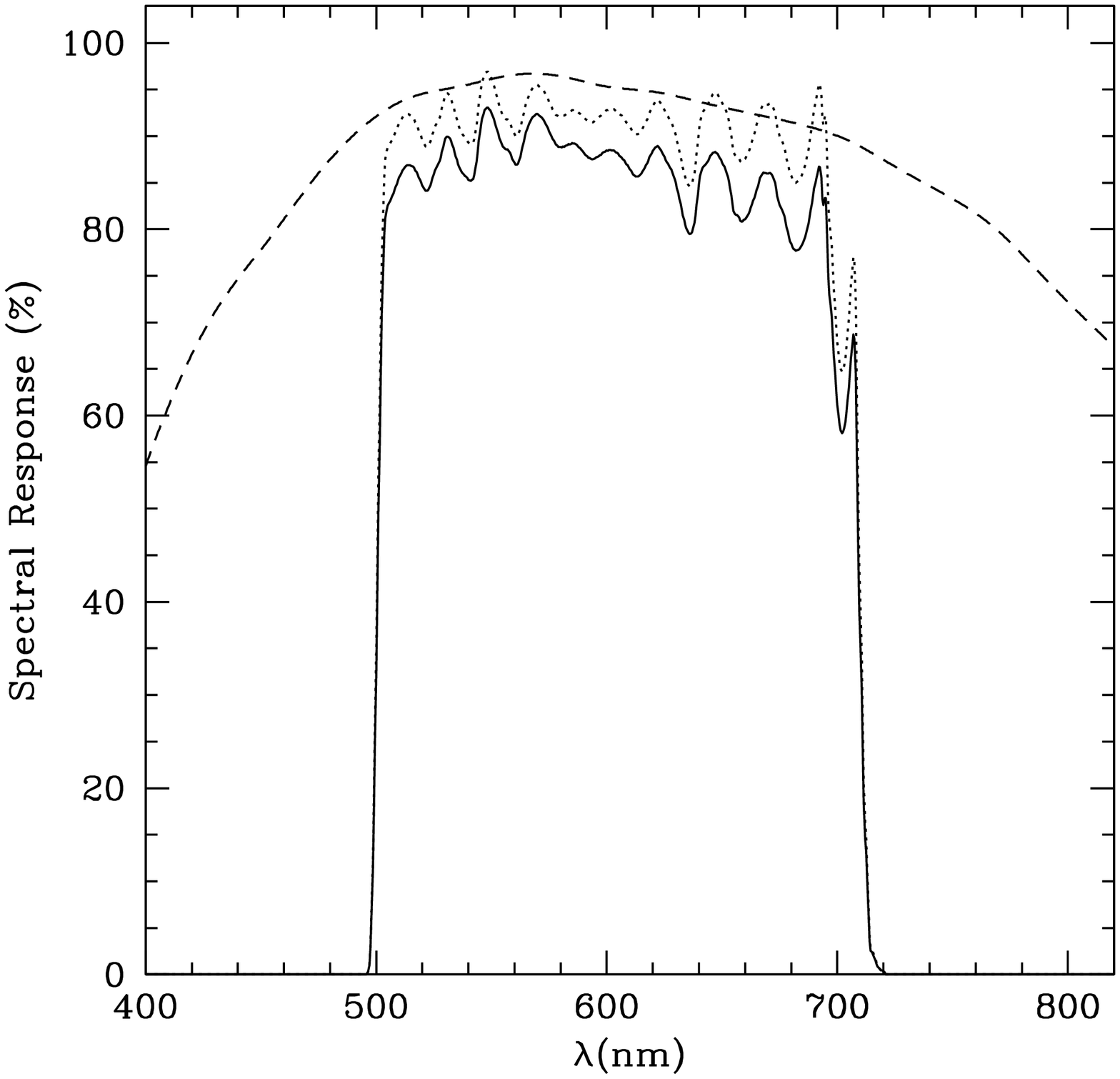}
\caption[]{Spectral response of the camera system. The dashed line
  shows the quantum efficiency of the CCD, the dotted line shows the
  filter transmission, and the solid line shows the total spectral
  response. The quantum efficiency data for the CCD was obtained from
  the e2v website at \url{http://www.e2v.com}.}
\label{fig:qe}
\end{figure}

Each telescope is equipped with a thermoelectrically cooled Spectral
Instruments Series 800 CCD camera employing a thinned,
backside-illuminated e2v $2048 \times 2052$ CCD42-40
chip.\footnote{These chips are specified as 2048$\times$2048, but
  there are four extra rows at the end of the CCD to act as a buffer
  for current leakage from the substrate. These rows must be read out
  in zipper mode as they are shifted with every other row. However,
  the current leakage is negligible at our 5~Hz image cadence, so we
  can use these extra rows as actual imaging area.} The CCDs have
13.5~$\mu$m pixels, corresponding to a plate scale of 2.9~arc~sec per
pixel, and the total image area is 27.6~mm~$\times$~27.6~mm. There are
two read channels, which work at a combined speed of
$1.8\times10^6$~pixels per second. The gain of each amplifier is set
to about 2.0~ADU/e$^-$, and the read noise is about 10~e$^-$. The
cameras are operated at a temperature of -20$^\circ$~C, and the dark
current is about 0.05~e$^-$/sec/pixel. Each CCD chip has the e2v
mid-band coating, providing peak quantum efficiency of more than 93\%
and mean quantum efficiency over the spectral range we will employ of
over 90\%. Each camera has a custom-made filter (produced by Custom
Scientific, Inc.)  which provides more than 80\% transmission between
\hbox{500 nm $\lesssim \lambda \lesssim$ 700 nm.}  The quantum
efficiency of the CCD and filter transmission curves are shown in
\fig{fig:qe}.

The camera was delivered with a chiller for the thermoelectric cooler,
but we found this difficult to operate remotely. Since we do not need
chilled water for our application the chillers were replaced with ITW
Cooling Systems 2500SS welding coolers, which are simply water
circulators with fan-cooled radiators. Each cooler is equipped with a
GEMS~2659 flow switch relay, which will signal if the coolant flow
stops for some reason.

\subsection{Weather Monitoring}
\label{sec:weather}
Accurate weather monitoring is an important component in any
robotic telescope system, especially at a site prone to rain and high
humidity. Most of the weather information is monitored by two Vaisala
WXT510 weather stations. Each weather station is controlled and read
through a serial port interface by a small Linux workstation. The two
systems are used for redundancy in case one of them fails. These
weather stations provide information such as temperature, humidity,
dew point, barometric pressure, and wind speed and direction.

To provide accurate measurement of the dew point under high humidity
conditions, we have installed a pair of Vaisala HMT337 humidity and
temperature sensors. These devices are designed for high humidity
applications, and specify an accuracy of $\pm 1.7$\% in relative
humidity (or more importantly, $\pm 0.3^\circ$~C accuracy in the dew
point). These devices are run by the same computers that control the
weather stations.

Each telescope enclosure is also equipped with a Vaisala DRD11A Rain
Detector to check for any precipitation. The sensors output an analog
voltage proportional to the amount of moisture on the sensor surface,
and these voltages are monitored by a computer in each enclosure by a
serial analog to digital converter. The devices are extremely
sensitive, and will trigger the closure of our enclosure lids
immediately after the detection of a single droplet.  Temperatures of
the various parts of each optical assembly (primary mirror, secondary
mirror, support struts, etc.) are monitored by Picotech Technology
Ltd. thermocouple sensors.

The weather parameters of most concern are precipitation, humidity,
wind, and temperature. We would like the lids closed if there is any
precipitation or if the humidity is high enough for condensation to
form on the telescope components. Wind is primarily a concern because
we do not want the lids open if the wind speed is high enough that the
telescopes might get damaged, or if we are not able to close the lid
at all (note that the lid effectively becomes a large sail during the
opening and closing operations due to the clamshell design, described
in \sect{sec:enclosures}). Wind is also a concern because of the
effect on the tracking accuracy of the telescopes. As discussed in
\sect{sec:zipper}, oscillations in the tracking accuracy due to wind
will decrease the signal-to-noise ratio achievable by the
system. Temperature monitoring is necessary because many of the
telescope components (e.g. motor controllers) are not rated below
certain temperatures. Temperatures of the optical and telescope
assembly are compared to the dew point temperature so we can close our
enclosure lids if the telescopes get cold enough for condensation to
form, as well as to provide real time adjustments to our focus
positions as the temperature changes. (This will be discussed in
detail in \sect{sec:focus}.)

\begin{deluxetable}{lll}
\tablecolumns{3}
\tablewidth{0pc}
\tablecaption{Weather alarm parameters}
\tablehead{Parameter & Alarm On Threshold & Alarm Off Threshold }
\startdata
DRD11A Voltage & $<$ 2.5 V & $>$ 2.8 V\\
Relative Humidity\tablenotemark{a} & $>$ 95\% & $<$ 94\%\\
Dew point -- temperature differential\tablenotemark{b} & $<$ 1.0$^\circ$~C &
$>$ 1.5$^\circ$~C\\ 
Average wind speed & $>$ 20 km/hr & $<$ 18 km/hr\\
Maximum wind speed\tablenotemark{c} & $>$ 30 km/hr & $<$ 28 km/hr\\
Minimum temperature & $< $-5.0$^\circ$~C & $>$ -3.0$^\circ$~C
\enddata
\tablenotetext{a}{Value reported by HMT337 dew point sensor.}
\tablenotetext{b}{Dew point value reported by HMT337 dew point sensor,
temperature differentials calculated from HMT337 temperature reading and
Picotech thermocouple measurements of telescope component temperatures.}
\tablenotetext{c}{Maximum wind speed measured over a 10~minute interval.}
\label{tab:weath}
\end{deluxetable}

We have therefore defined a set of alarm thresholds for these various
weather parameters. If one of these parameters crosses a threshold, a
``WEATH\_BAD'' alarm state is entered. If this alarm state is entered,
the lids of all of the enclosures will immediately be closed
automatically. In addition, we have defined a second set of thresholds
which need to be crossed for the WEATH\_BAD alarm state to be turned
off. If a WEATH\_BAD alarm state is entered, the weather parameters
must be past the threshold values for 3~minutes before the alarm is
turned off. This prevents the alarm from rapidly turning on and off
when one of the parameters is hovering near the threshold. These
thresholds are shown in \tbl{tab:weath}.

\begin{figure}[bt]
\plotone{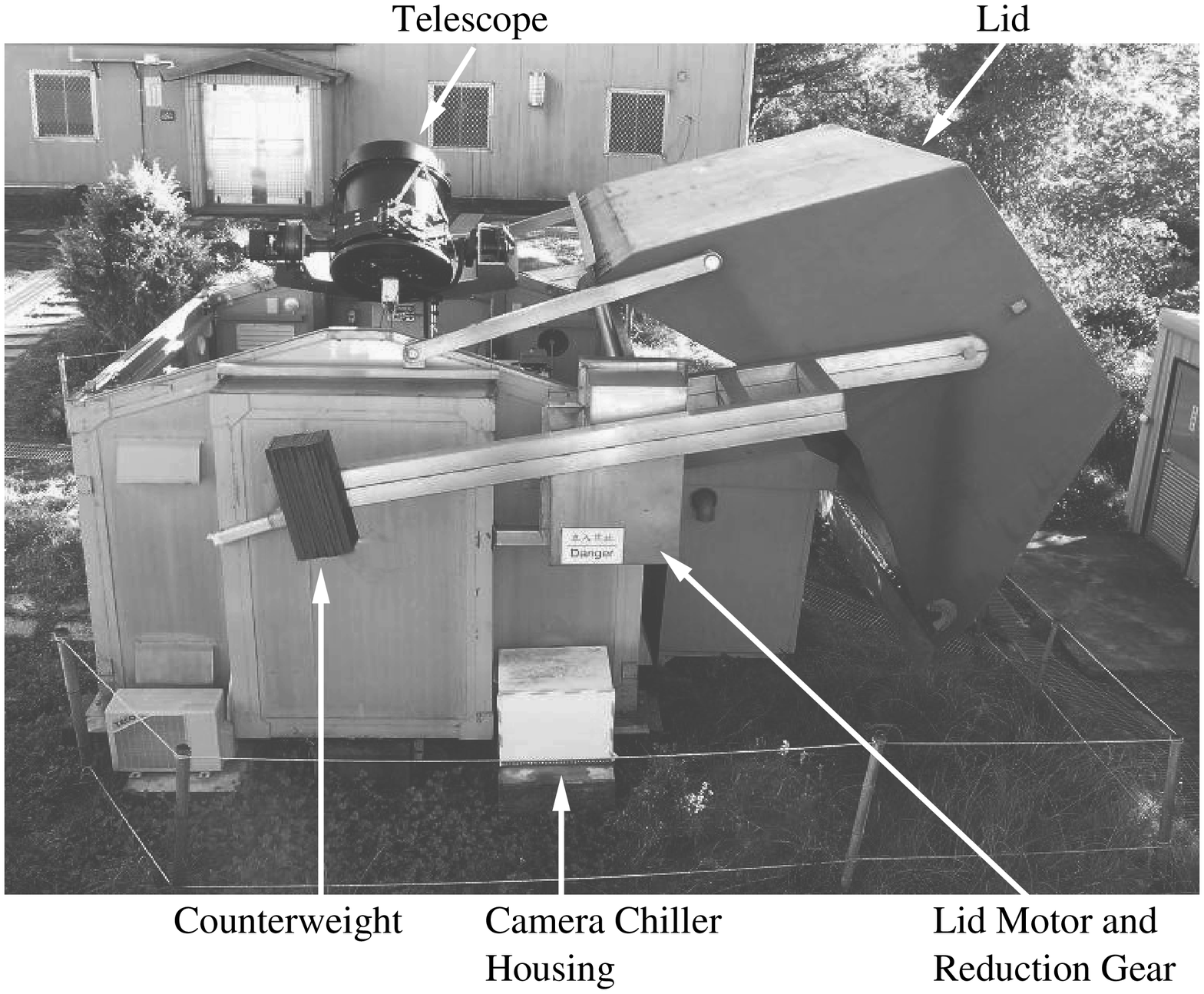}
\caption[]{The enclosure for telescope TAOS~A.}
\label{fig:enclosure}
\end{figure}

\subsection{The Enclosures}
\label{sec:enclosures}
The TAOS enclosures (see \fig{fig:enclosure}) were designed to
withstand the hurricane force winds which accompany the four or five
typhoons which typically move through the region every summer. The
enclosure walls consist of two 1~mm thick steel plates separated by
5~cm of insulating foam and are supported by \hbox{10 cm $\times$ 10
  cm} steel beams.  The working area is about \hbox{3 m $\times$ 3
  m}. The lids are of a counterweighted clamshell type design. They
are constructed of polyurethane-filled fiberglass and were fabricated
by the Aeronautical Research Laboratory of the Chungshan Institute of
Science and Technology in Taiwan. Each lid is actuated by an ADLEE
BM-370 brushless DC motor and a 1:400 worm reducer made by Shin Wei
Ent Co.  A custom control circuit is used to provide both manual
operation and automated robotic operation through software. The time
to open or close the lid is about one minute. In order to keep the lid
operational during a power failure, the motor is powered by a SMR
MCS1800 DC UPS made by Delta Electronics Inc.

Each enclosure is equipped with air-conditioning to keep the
temperature of the telescope close to the anticipated night time
temperature.  This improves image quality by minimizing the
temperature difference between the telescope components and the
ambient temperature, and also minimizes the time used to find the best
focus and focus change during the night.

Finally, each enclosure is equipped with a custom isolation
transformer to protect the hardware from power surges induced by
lightning. All four TAOS enclosures are connected to the control
building by 1~Gbps fiber optic cable for lightning protection as well.

\subsection{The Robotic Observatory Control System}
\label{sec:iobox}
In each enclosure, every subsystem other than the telescope and camera
is controlled by a Sealevel Systems Model 8010 32~bit I/O card which
is connected to a custom built interface box. These boxes are based on a
design used by the ROTSE Survey~\citep{2003PASP..115..132A}, but the
design has been extensively modified. The boxes control the power to
the cameras, the camera coolers, and the telescopes. The boxes also
monitor the limit switches on the telescopes, and will automatically
cut the power to a telescope if one of the switches is tripped. This
is done in hardware with a series of relay switches to ensure
reliability. The chiller flow switches are also monitored by these
systems, and the software will turn off the thermoelectric cooler on
the CCD and cut the power to the chiller should the flow unexpectedly
stop.

The lids are also controlled by these interface boxes. The lids are
opened and closed by asserting 12~V signal lines to the lid controller
boxes. The limit switches to the lids are also monitored, and once again
a set of relay switches will automatically cut the power to the
open/close line when the open/close limit switch is tripped.

Finally, each box is equipped with an ICS Advent WDT5 watchdog
module. This module must be pinged by the I/O card once every second,
or it will close a set of relay switches which will automatically cut
the power to the telescope and close the lid. If the computer crashes
or one of the control daemons should happen to die, the system will
then be safe from any inclement weather which should arise before the
system can be restarted.

\section{High-Speed Photometry with Zipper Mode Readout}
\label{sec:zipper}
\begin{figure}[bt]
\plotone{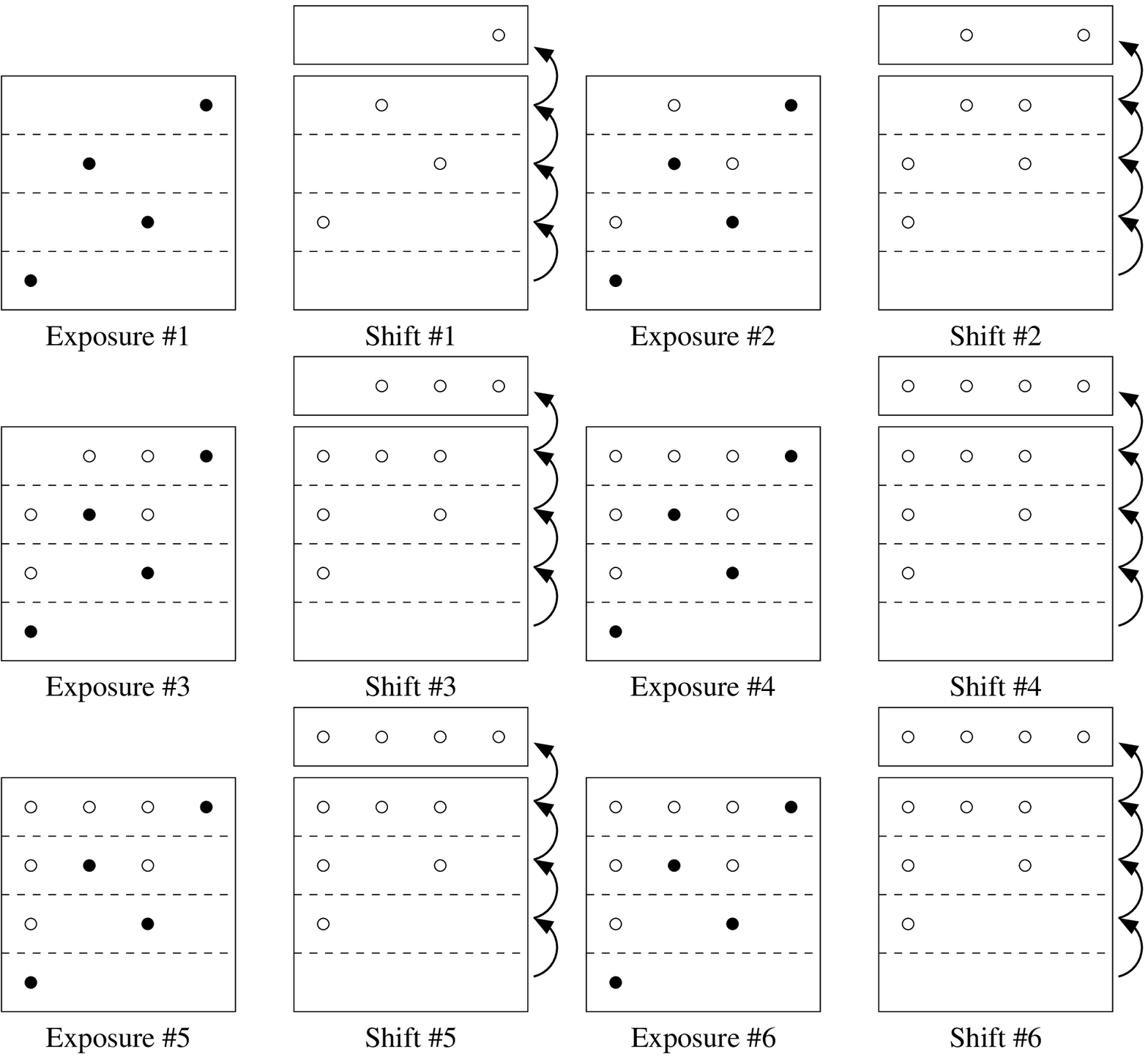}
\caption[]{Illustration of zipper mode operation with the focal plane
  divided into four row blocks. Solid circles indicate actual stars,
  empty circles indicate locations of electrons after shifts. Note
  that after the fourth sequence of expose and shift operations,
  flux from all four stars is read out at each subsequent shift,
  although the flux from these stars was collected at different
  epochs.}
\label{fig:zip}
\end{figure}

The CCDs used for the TAOS project are described in \sect{sec:camera}.
The devices are 2048$\times$2052 e2v CCD42-40 chips. Each chip has two
amplifiers, which allow for a combined readout rate of 1.8~MHz. It
takes approximately 2.3~seconds to read out the entire frame, which
makes our desired 5~Hz image cadence impossible without more than 90\%
dead time. To overcome this limitation, we have designed a novel CCD
readout method, which we call \emph{zipper mode} (see \fig{fig:zip}).

\begin{figure}
\plotone{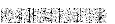}
\caption[]{
  Example of a zipper mode image comprising a $512\times76$
  subsection of a $2048\times76$ row block.}
\label{fig:rowblock}
\end{figure}

As with conventional operation, when reading out in zipper mode, the
telescope tracks the sky and stars are imaged onto the focal
plane. However, instead of simply opening and closing the shutter for
each exposure, the shutter is left open, and after 105~ms, a ``row
block'' of 76~rows is read out. (The read out operation takes
approximately 95~ms, so the readout time plus the exposure time is
equal to our required sampling time of 200~ms.) This is followed by
another ``hold'' of 105~ms, then another row block is read out. This
cycle may be repeated for extended periods of time, possibly up to
several hours.  Note that after 27~row blocks are read out (27~row
blocks~$\times$~76~rows~$=$~2052~rows, the total number of rows in the
CCD), photoelectrons from every star on the focal plane will be read
out in each subsequent row block.

The ability provided by zipper mode to read at 5~Hz comes with a cost
in signal-to-noise. This is due to two properties inherent to zipper
mode, and any attempt to alleviate one of the problems exacerbates the
other. First, photons from sky background are collected continuously
during each hold and shift, while photons from a star are collected
only during a single hold. Note that with row blocks of 76~rows, the
pixels in a single row block contain photoelectrons collected from a
total of $2052 / 76 = 27$~holds. With a hold time of 105~ms and a
sampling cadence of 5~Hz, the average recorded number of sky
photoelectrons per pixel is thus a factor of \hbox{$27 \times 200 /
  105 = 51.4$}~times brighter than it would be after a simple, 105~ms
exposure. Second, it takes just over 1~ms to read a single row.
Photons from stars are still collected during this time, meaning that
stars will leave streaks as a row block is read out, as can be seen in
\fig{fig:rowblock}. Faint stars are particularly vulnerable to this
effect of flux contamination by streaks of bright stars. One could
conceivably reduce the sky background using row blocks with a larger
number of rows, but this would reduce the effective exposure time
relative to the readout time and reduce the measured stellar
brightness relative to the brightnesses of the streaks.

We would ideally like the row block size to be an integral factor of
2052, the total number of rows on the CCD. (Stars near the edges of an
image are missing a fraction of their flux, increasing uncertainty on
the photometry not only of the edge stars but also of their neighbors.
This effect is minimized if the row block size is an integral factor
of the number of rows in the CCD.)  We found the signal-to-noise ratio
was highest when the read time was close to the exposure time, and the
integral factor of 2052 meeting this requirement is 76~rows.

Because of the increased sky background, zipper mode works well only
for relatively bright stars. On a dark night, adequate signal-to-noise
($\sim$7) can be achieved with a star of $V \sim\! 13.5$, which
consequently is the faintest magnitude we use in our occultation
search.

\begin{figure}
\plotone{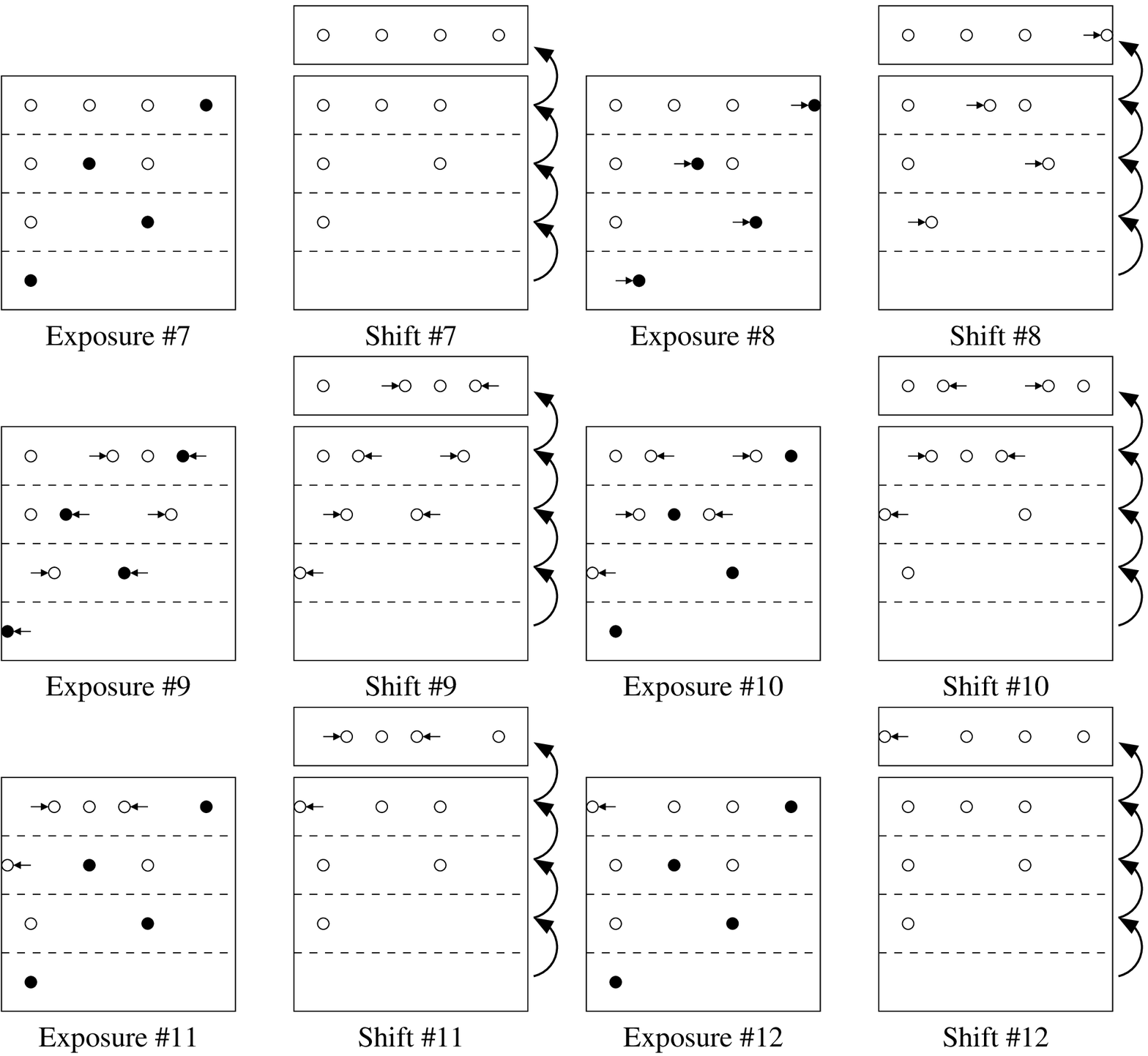}
\caption[]{Continuation of zipper mode schematic shown in
  Figure~\protect\ref{fig:zip}, illustrating effect of tracking errors
  throughout a zipper mode run. In this illustration, the telescope
  drifts to the left during exposure \#8, moving the stars to the
  right of the image. During exposure \#9, the telescope moves too far
  back to the right, moving the stars to the left of the image. The
  arrows on the images indicate the displacements of the stars from
  their nominal positions. Note that the stellar flux that is read out
  in a row block is collected at different times, depending on where a
  star is located on the focal plane. Neighboring stars on a row block
  could therefore have offsets from their nominal positions in
  different directions.}
\label{fig:zipshift}
\end{figure}

This readout mode also requires accurate pointing and tracking
throughout the duration of a zipper mode run, ideally to within a
small fraction ($\sim 0.1$) of a pixel. As shown in
\fig{fig:zipshift}, if there are significant oscillations in pointing
error, then neighboring stars in a row block can move in different
directions from their nominal positions over time. This can cause a
significant reduction in signal-to-noise for closely neighboring stars
in a row block due to varying blending fractions over time.

\begin{figure}[bt]
\plotone{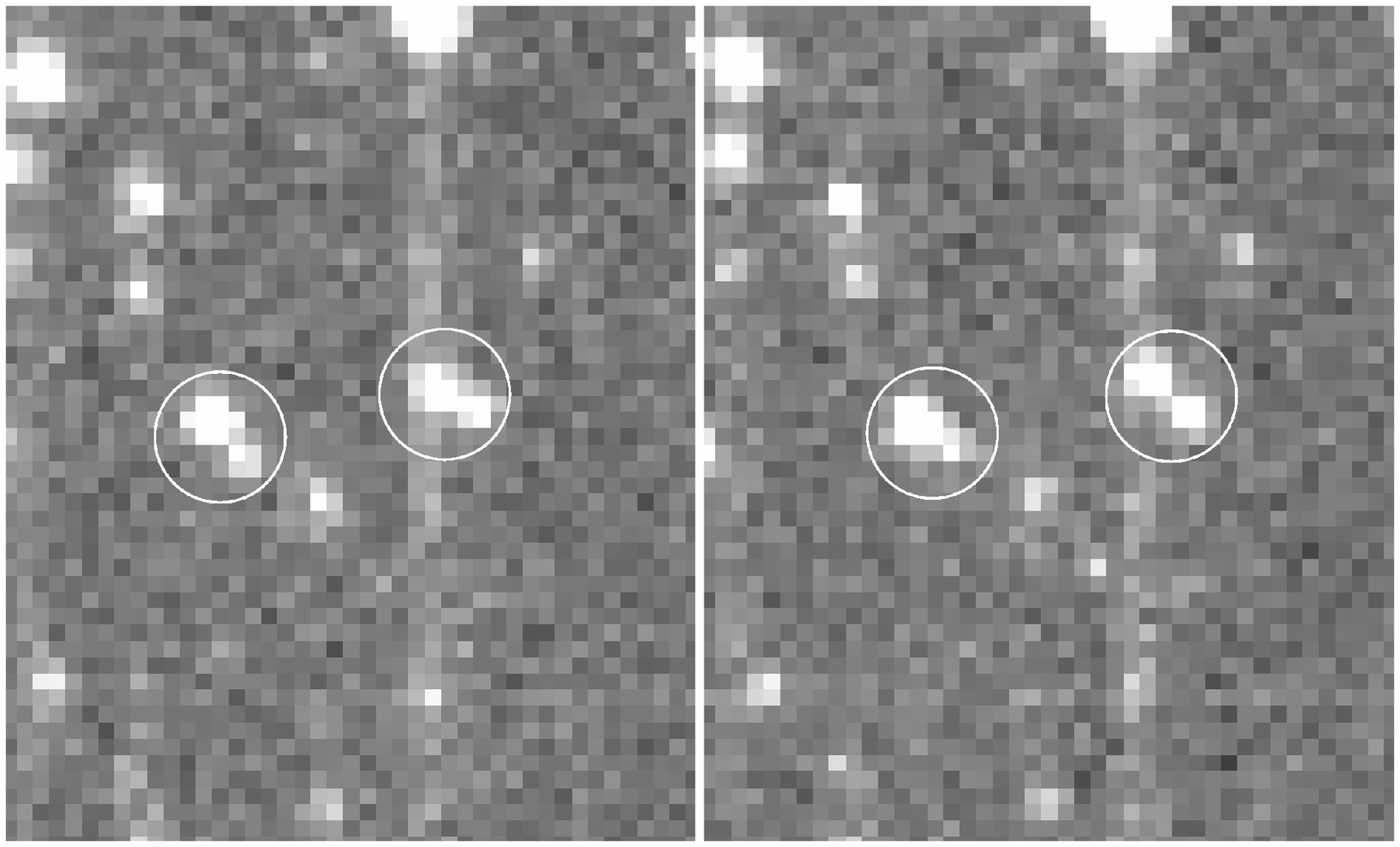}
\caption[]{Two row block subsections taken at two different
  epochs. Note the two pairs of stars inside the circles. The two
  neighboring stars in the right circle are close together in the left
  image and have moved apart in the right (later) image. The two stars
  in the left circle moved closer together in the right image.}
\label{fig:neighbor}
\end{figure}

We did find both significant drifts and oscillations in the tracking
of the telescopes, as shown in \fig{fig:neighbor}. Reducing the
magnitudes of these errors required significant hardware improvements
(see \sect{sec:telescopes}) and software modifications to the
telescopes (see \sect{sec:teld}). These modifications were completed
in September of 2005, so a large fraction of our first dataset
suffered from this problem. This will be discussed in
\citet{kiwiphot}.

\section{The Software}

\begin{figure}[bt]
\epsscale{.8}
\plotone{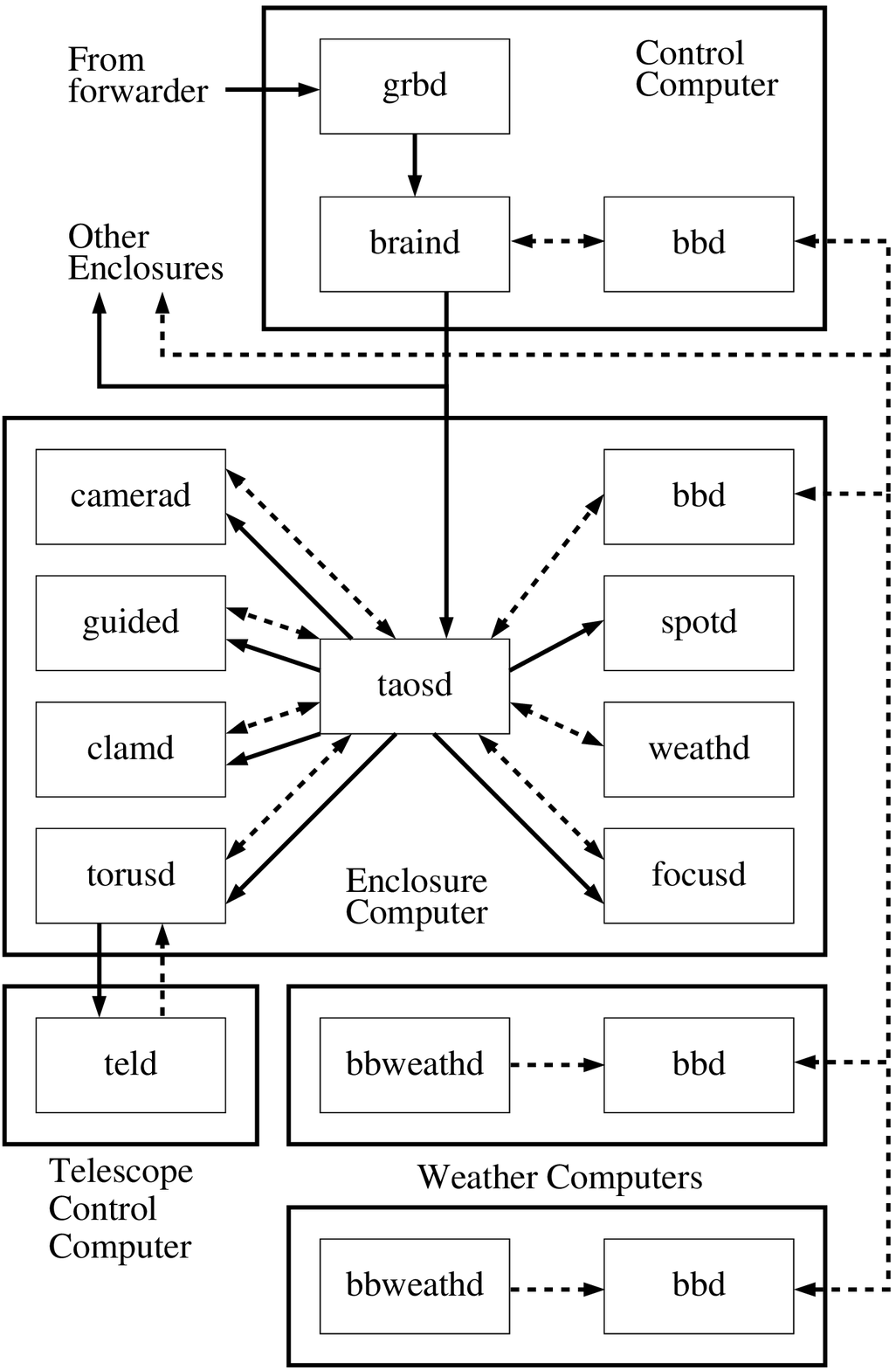}
\caption[]{Schematic of the TAOS system control software. Solid arrows
  indicate control message bus, and dashed arrows indicate status
  message bus. See text for details.}
\label{fig:software}
\end{figure}

The control software for the TAOS system is based on the control
software used for the ROTSE Survey \citep{2003PASP..115..132A},
however it has been extensively modified for TAOS. Each component of
the system is controlled by a separate daemon. A schematic of the
daemons and their communication paths can be seen in
\fig{fig:software}, and a description of each daemon is given in the
following subsections.

Each enclosure has a dual-CPU Pentium~4 computer to control all of the
hardware described in the previous section, and a small PC104 computer
to control the telescope via the OMS PC68 Motion Controller Card. The
main control computer is located in the TAOS control building, which
also houses two small computers to monitor the two sets of weather
stations (see \sect{sec:weather}). All of the computers are running
the Debian Sarge\footnote{\url{http://www.debian.org}} distribution of
Linux. Each of the main enclosure computers are running Linux kernel
version 2.4.18 with the Kansas University Real-Time patch
(KURT\footnote{\url{http://www.ittc.ku.edu/kurt/}}) to provide
microsecond timing resolution, which is important for synchronized
zipper mode imaging at 5~Hz. The main control computer, weather
station computers and telescope control computers are running kernel
version 2.6.8.

In order to provide accurate timing to the control computers
(necessary for accurate telescope pointing and synchronized high-speed
imaging), the clocks of all of the computers in the control system are
synchronized via \emph{ntp}\footnote{Network Time Protocol. See
  \url{http://www.ntp.org/}.} to the control computer. The ntp server
on control computer synchronizes to two ntp servers in
Taiwan\footnote{tick.stdtime.gov.tw, stdtime.sinica.edu.tw}.

\subsection{Command and Status Message Passing}
All command and status information is passed from computer to computer
via standard \emph{inet} sockets using a custom message passing
library. Command information is sent from the control computer
directly to each enclosure computer.

Each computer has a System V shared memory area which contains all
status information from each enclosure, the control computer, and the
weather computers. We call the shared memory area the \emph{bulletin
  board}, and the status information is kept up to date by the
bulletin board daemon, \bbd. There is an instance of \bbd\ running on
each computer in the local network, and once every second each
instance of \bbd\ updates every other instance with its local status
information. Each daemon on the system thus has available up-to-date
information on every other daemon running on the system.

\begin{deluxetable}{lcccc}
\tablecolumns{3}
\tablewidth{0pc}
\tablecaption{Scheduler states}
\tablehead{State & Sun Elevation & Weather Status & Lid Position & Telescope and Camera Power}
\startdata
SHUTDOWN & $\ge 0^\circ$ & good or bad & closed & off\\
GRBMODE & $< 0^\circ$ & good & open & on\\
STARTUP & $< -18^\circ$ & good & open & on\\
STANDBY & $< 0^\circ$ & bad & closed & on\\
\enddata
\label{tab:braindstate}
\end{deluxetable}

\subsection{The Scheduler Daemon \braind}
The scheduler has four internal states, described in
\tbl{tab:braindstate}.  While the Sun is above the horizon, the
scheduler is in the SHUTDOWN state. In this state the lid is closed
and the telescope and camera are powered off. When the Sun goes below
the horizon and the weather is good (see \sect{sec:weather}), the
scheduler will enter the GRBMODE state. In this state, the lid will be
opened and the telescope and camera will be powered on. No zipper mode
occultation survey data will be collected in this state, however the
system will respond to any GRB alerts that come in. (Zipper mode data
are not so useful in twilight conditions due to the increase in sky
background, however GRB follow-up data could be useful under this
circumstance.) When the Sun goes below an elevation of $-18^\circ$
(i.e. the sky is sufficiently dark), \braind\ enters STARTUP mode, and
it will subsequently commence zipper mode observations for the
occultation survey. If any GRB alerts come in during a zipper mode
run, \braind\ will interrupt the observations and schedule follow-up
observations for the GRB event.  At the end of the evening, when the
Sun rises above $-18^\circ$, \braind\ will go back into GRBMODE and
will then only respond to GRB alerts. At sunrise, \braind\ will go
back into SHUTDOWN mode for the day, and will close the lids and power
down the telescopes and cameras.

If \braind\ is in the GRBMODE or STARTUP states and at any time the
weather should turn bad, the system will go into STANDBY mode and
close the lids. If the weather becomes good again, \braind\ will go
into either the GRBMODE or STARTUP state, depending on the elevation
of the Sun. If the weather is bad when the Sun sets, the system will
go straight into the STANDBY state from the SHUTDOWN
state. Conversely, if the system is in the STANDBY state at the end of
the night when the sun rises, it will go into the SHUTDOWN state.

Every evening after sunrise and sunset, \braind\ will schedule a set
of dark images to be taken on each camera. At sunset, this happens
300~seconds after \braind\ enters the GRBMODE state to give the camera
ample time to cool down. At sunrise, the dark images are taken after
the lids are closed but before the system enters the SHUTDOWN
state. Each series of dark images comprises a set of 20~stare mode dark
images with 1~second exposures, and a series of 10~zipper mode dark
images with the standard zipper mode parameters (5~Hz sampling, blocks
of 76~rows, and 32~row blocks per FITS file).

\subsection{The Weather Daemon \bbweathd}
The weather stations and dew point sensors described in
\sect{sec:weather} are read by two instances of \bbweathd, one
instance for each set of one weather station and one dew point
sensor. This daemon (as the name implies) simply reads both devices
and publishes the information on the bulletin board for other software
components to monitor. In addition, \bbweathd\ will check the measured
weather parameters against a set of limits, and set an alarm if
adverse weather is detected. This alarm state is also written to the
bulletin board for other daemons to monitor.

\subsection{Control Daemons}
In addition to the instance of \bbd\ described above, each
enclosure computer runs a set of eight control daemons, each dedicated
to a particular hardware component (or set of components) or high
level control task. These daemons are described in the following
subsections.

\subsubsection{The Master Control Daemon \taosd}
Every daemon running on the enclosure computer is a child of
\taosd. This daemon is responsible for forwarding commands from
\braind\ to the appropriate child daemon, and for monitoring the status
of each of the children and passing the status information on to the
local instance of \bbd. Each child daemon has a dedicated System~V
shared memory area which is used to pass command and status
information to and from \taosd. This daemon will automatically send
commands to certain daemons based on the status information of other
daemons (e.g. it will close the lid if adverse weather is detected).

\subsubsection{The Watchdog Daemon \spotd}
This simple daemon is responsible for pinging the watchdog module in
the observatory control interface box described in \sect{sec:iobox}.
The daemon is designed to run independently of the rest of the daemons
on the system to enable manual control (e.g. so the lid can be opened
manually without the rest of the software running), however when
\taosd\ is running, a simple ``I'M ALIVE'' command is expected from
\taosd\ at a rate of 1~Hz. If this command times out, \spotd\ will
die, the telescope will be stopped and the lid will automatically be
closed. Thus if \taosd\ should crash or otherwise be hung up for some
reason, or if the enclosure computer itself should crash, the system
will automatically go into a safe mode.

\subsubsection{The Local Weather Daemon \weathd}
While general weather information is monitored by \bbweathd\ as
described above, a local weather daemon \weathd\ runs on each
enclosure computer as well. This daemon has three tasks. First,
\weathd\ monitors the Vaisala precipitation sensors described in
\sect{sec:weather}. Second, it monitors the local copy of the
\bbweathd\ information on the bulletin board. If precipitation is
detected or \bbweathd\ reports adverse weather, a flag is set in the
\weathd\ shared memory segment which will instruct \taosd\ to abort
any current observations and close the lids.  Finally,
\weathd\ monitors the telescope strut and mirror temperatures read by
the thermocouple sensors attached to the telescope (see
\sect{sec:telescopes}), and compares these temperatures to the dew
point reported by \bbweathd\ to see if the lid needs to be closed to
prevent condensation on the telescope.

\subsubsection{The Telescope Command and Status Daemon \torusd}
The telescope is directly controlled by a small PC104 computer via an
OMS PC68 motion controller card, and communication between the main
enclosure computer and the PC104 computer is handled by \torusd. This
daemon connects to the telescope control daemon \teld\ (see
\sect{sec:teld}) over a set of four inet sockets, one for pointing and
tracking control, one for the focus control, one for status
information and one for miscellaneous control commands.

\subsubsection{The Guide Daemon \guided}
\label{sec:guided}
While we have greatly improved the pointing and tracking capabilities
of the system through both software and hardware modifications, we
found the pointing was still inadequate for the survey. Accurate
pointing (within 0.3~arc~sec) must be maintained throughout an entire
zipper mode run, which can last as long as 2~hours. As discussed in
\sect{sec:zipper}, poor tracking can significantly decrease the
signal-to-noise performance of the system, so it is desirable to keep
the tracking as accurate as possible. While many surveys have a
separate guiding imager, TAOS has the advantage that images are
collected at a high rate, and we use this feature to enable our
guiding system.

The guide daemon runs in two modes. During the initial pointing phase,
we point the telescope to a field center and take a stare mode
image. This image is analyzed to find the actual RA and Dec of the
image center, and a pointing offset is calculated. A pointing
adjustment command is sent to \torusd, and the process is repeated
until the center of the image is within 0.1~pixels of the actual field
center. In this way, at the beginning of a zipper run, all four
telescopes are pointing to within 0.3~arc~sec of the target field
center.

The image analysis routine makes use of the standard software packages
\emph{SExtractor} \citep{sextractor} and \emph{wcstools}
\citep{2006ASPC..351..204M} used in conjunction with the USNO-B1.0
catalog \citep{2003AJ....125..984M}. First, after an image is taken,
the \emph{wcstools} utility \emph{imwcs} is run to find the World
Coordinate System (WCS) keywords to transform RA and Dec to pixel
coordinates. Second, \emph{SExtractor} is run to find objects in the
image. Third, the \emph{SExtractor} output is used with the
\emph{wcstools} utility \emph{immatch} to match the objects found by
\emph{SExtractor}. Finally, a set of polynomials
\citep{1997hstc.work..144M} are fit to the data to convert from $x$
and $y$, the pixel coordinates of each matched star calculated from
the WCS coefficients, to $x_p$ and $y_p$, the actual centroid
locations reported by \emph{SExtractor}.  The polynomials are of the
form
\begin{eqnarray}
x_p & = & C_0 + C_1 x + C_2 y + C_3 x^2 + C_4 xy + C_5 y^2 \nonumber\\
& & + C_6 x^3 + C_7 x^2 y + C_8 xy^2 + C_9 y^3 \nonumber\\
y_p & = & D_0 + D_1 x + D_2 y + D_3 x^2 + D_4 xy + D_5 y^2 \nonumber\\
& & + D_6 x^3 + D_7 x^2 y + D_8 xy^2 + D_9 y^3, \nonumber
\end{eqnarray}
where $C_i$ and $D_i$ are the fit coefficients. After the fits are
complete, these distortion coefficients and WCS keywords are used to
calculate the RA and Dec of the center of the image, which is then used to
calculate the pointing offset. The pointing offset is written to
the \guided\ shared memory status segment, which is then read
by \taosd\ and passed to \torusd\ as a pointing adjustment command.

The second \guided\ mode is the tracking mode. At the end of the
initial pointing phase, 20~guide stars are selected from the stare
mode image. The stars are selected on criteria including moderate
brightness (i.e. bright enough to have good signal-to-noise
characteristics but not so bright as to leave significant streaks in
the zipper images), isolated (no close neighbors in zipper mode
images), and they should be spread out through the image so the entire
focal plane is well sampled. Note that a set of 32~consecutive row
blocks is typically written to each FITS file, so there are
effectively 640~guide stars in each image. Every 30~seconds, the last
FITS image that was written to disk is analyzed. \emph{SExtractor} is
run on the image, and the centroids of the guide stars are found and
compared with the original values. A pointing correction is calculated
and written to the \guided\ shared memory status segment, which in
turn is read by \taosd\ and passed as a pointing adjustment command to
\torusd. We find that the inherent tracking of the system is
sufficient that the pointing drifts by much less than 0.1~pixel in the
30~second guiding update period, and \guided\ is thus able to keep the
pointing correct throughout an entire zipper mode run of up to two
hours in duration.

The guide daemon also monitors image rotation throughout the course of
a zipper run. Taiwan suffers from frequent earthquakes which can
disrupt the polar alignment of the telescopes. We can thus learn in
real time if the telescopes need realignment. Furthermore,
\guided\ reports the visibility conditions throughout the duration of
a zipper mode run by monitoring the star count.  Each time a zipper
mode FITS file is analyzed for tracking adjustments, \emph{SExtractor}
reports the number of stars to which it can fit a standard PSF. The
star count is compared to the number of stars found at the beginning
of the zipper mode run to indicate how the observing conditions vary
throughout the run. In our fully automated observing mode,
\taosd\ will interrupt observations if the star count drops
significantly. This usually occurs due to incoming clouds, or if the
seeing degenerates due to humidity or focusing problems. This
real-time image quality analysis helps optimize our data collection as
we can interrupt a zipper mode run and refocus the telescope if
necessary.

\subsubsection{The Lid Control Daemon \clamd}
The lid is opened and closed by a relatively simple daemon
\clamd. This daemon will assert an open or close signal to the lid
controller on command, and monitor the open and close limit switches
on the lid. The open and close signals are lowered when the
appropriate limit switch is hit. The daemon will also set an alarm if
the open or close operation times out, and this alarm will alert
someone to fix the problem manually.

\subsubsection{The Focus Control Daemon \focusd}
\label{sec:focus}
The focus control daemon \focusd\ has two tasks. The first is to
analyze images taken during a \emph{focus sequence} to determine the
ideal focus position. A focus sequence consists of a series of
one~second exposures taken with secondary mirror at different focus
positions. A total of 17 images are acquired while moving the focus
position by 15~$\mu$m steps. A 1k$\times$1k sub-image from the center
of the focal plane is analyzed using \emph{SExtractor} to find the
average Full Width Half Maximum (FWHM) of each star in the image, and
the focus position and medium FWHM are logged internally. After all of
the images have been collected and analyzed, the FWHM data are fit to
two straight lines, one line on each side of the point with the
minimum FWHM. These fits are then used to calculate the best focus
position, which is then entered into a database along with the
telescope component temperatures (reported by \weathd) and zenith
angle of the telescope (reported by \torusd).

The second function of the focus daemon is to monitor the telescope
component temperatures and zenith angle throughout a zipper mode run
and to look up the ideal focus position in the database. If the ideal
focus position changes, the new position is written to the \focusd\
shared memory status segment, where it is read by \taosd\ and
passed on to \torusd\ as a focus adjustment command.

\subsection{The Telescope Control Daemon \teld}
\label{sec:teld}

\begin{deluxetable}{ll}
\tablecolumns{2}
\tablewidth{0pc}
\tablecaption{Pointing model coefficients}
\tablehead{Coefficient & Description}
\startdata
IH & Hour angle home switch position \\
ID & Declination angle home switch position \\
CH & Collimation error\\
NP & Non-perpendicularity of $h$ and $\delta$ axes\\
MA & Azimuthal polar alignment error\\
ME & Elevation polar alignment error\\
TF & Tube flexure\\
FO & Fork flexure\\
KZ and MZ & First order correction to declination encoder error\\
\enddata
\label{tab:pm}
\end{deluxetable}

The system originally used the \emph{telescoped} program from the
OCAAS\footnote{Observatory Control and Astronomical Analysis System,
  see \url{http://www.clearskyinstitute.com/}} software package to
control the telescope. However, we found three problems with this code
that led us to the decision to write a completely new software
package. First, the pointing model used by OCAAS did not accurately
represent the pointing actually realized with the system. The
residuals between the model and the true pointing were often as high
as 10 arc~minutes. Second, we found it difficult to generate the
pointing model under OCAAS, where the pointing is adjusted by hand
until a catalog star is centered on the image, and this is repeated
for several stars. This was very time-consuming as it needs to be done
several times per year (after earthquakes), and it needs to be done
for all four telescopes. It is also difficult to do remotely. Third,
we found it difficult to implement a robust communication interface
between \torusd\ and the OCAAS software without a significant rewrite
of the software.

The replacement daemon, \teld\, is a multi-threaded program with
separate threads dedicated to telescope pointing, focus control, status
reporting, and master control (i.e. a parent thread to monitor the
children). In addition, when tracking a field a new thread is spawned
to control the telescope tracking directly. This thread runs with an
elevated scheduling priority to keep the timing of the control loop
accurate. Implementing a multi-threaded solution to the telescope
control also required rewriting the driver for the PC68 motion
controller card to make it thread safe. This was done by simply adding
a semaphore to the driver which is raised whenever the card is
accessed.

The daemon makes use of the SLALIB Positional Astronomy
Library\footnote{\url{http://star-www.rl.ac.uk/}} to convert to and
from J2000 RA and Dec to observed HA and Dec ($\hobs$ and $\dobs$). A
custom pointing model derived from that used by the TPOINT software
package\footnote{\url{http://www.tpsoft.demon.co.uk/}} is used to
convert $\hobs$ and $\dobs$ to actual encoder angles ($\theta_h$ for
the hour angle encoder and $\theta_\delta$ for the declination
encoder). The pointing model used is:
\begin{eqnarray}
h_1 & = & \hobs - \mathrm{TF} \cos\tlat\sin\hobs/
\cos\dobs\nonumber\\
\delta_1 & = & \dobs  - \mathrm{TF} (\cos\tlat\cos\hobs\sin\dobs\nonumber\\
& &  - \sin\tlat\cos\dobs) \nonumber\\
& &  - \mathrm{KZ}\cos\dobs
 - \mathrm{MZ}\sin\dobs\nonumber\\
& & - \mathrm{FO} \cos\hobs\nonumber\\
\theta_h & = & -h_1 + \mathrm{CH} \sec \delta_1 +
\mathrm{NP}\tan \delta_1 + \nonumber\\
& & (\mathrm{ME}\sin h_1  - \mathrm{MA} \cos h_1) \tan \delta_1 +
\mathrm{IH} \nonumber\\
\theta_\delta & = & \delta_1 - \mathrm{ME}\cos h_1 - \mathrm{MA} \sin h_1 -
\mathrm{ID}\nonumber
\end{eqnarray} 
where $h_1$ and $\delta_1$ are temporary variables, $\tlat$ is the
geodetic latitude of the site and the ten pointing model coefficients
are described in Table~\ref{tab:pm}.

After implementing \teld, we found the typical residuals between the
calculated and true positions to be less than 10~arc~seconds. Used in
conjunction with \guided, we found that we can track to well within
0.3~arc~sec for durations up to at least two hours.

We also implemented an automatic pointing model generation feature to
the system. A set of 40~images is taken at a grid of locations on the
sky, and the images are analyzed by \guided\ to find the actual RA and
Dec of the centers of the images. The subsequent arrays of RA, Dec,
$\theta_h$, $\theta_\delta$, and epoch are fit using the \emph{amoeba}
fitting routine adapted from \citet{numrec}. This works much faster
than the OCAAS method, especially when it must be done to all four
telescopes several times a year. This also allows a new pointing model
to be created remotely and automatically.

\subsection{Collection of Zipper Mode Data}
When the system is in STARTUP mode, \braind\ will schedule a series of
zipper mode \emph{runs}. At the start of a run, \braind\ will choose a
standard TAOS field (see \sect{sec:fields}) based on criteria such as
distance from the ecliptic plane, distance from zenith, and number of
stars in the field. The scheduler will then instruct \torusd\ to point
all of the telescopes at the field center. Each telescope will then
undergo a focus sequence (see \sect{sec:focus}) and set the secondary
focus to the best position. Once this is completed, each telescope
will undergo a pointing correction, as described in
\sect{sec:guided}. Once all of the telescopes are pointing correctly,
a set of three 1~second stare mode images are acquired for diagnostic
purposes. Next, the zipper mode sequence is started. Each camera
system is sent a start time by \braind, and when the start time is
reached, the shutters are opened and zipper mode readout is
commenced. This 5~Hz readout process continues for a total of
90~minutes (as long as the weather remains good and no GRB alerts are
received). Throughout the duration of the run, \guided\ will keep the
telescope pointed correctly. At the end of the run, a second set of
three diagnostic stare mode images are acquired, also with 1~second
exposures. Finally, a sequence of 1~second stare mode images is
acquired with pointing offsets of 0.8$^\circ$ on a 3$\times$3 grid
about the center of the field. This provides an effective mosaic image
of $\sim$11$\Box^\circ$ centered on the field, which is also used for
diagnostic purposes.

Throughout a zipper mode run, \braind\ will monitor the results of
each command and the status of each daemon. If an error occurs,
\braind\ will attempt to recover by restarting any software that needs
it and, if necessary, reinitializing the zipper mode run. If
\guided\ reports bad observing conditions, the zipper run will be
aborted. A new field may be selected at this time based on the
scheduling criteria discussed above, and if so \braind\ will instruct
the telescopes to point to the new field. In any case, the system will
then undergo a new focus sequence to refocus the telescopes. If the
weather is cloudy, this step will fail due to the lack of stars in the
images. In this case, the system will wait for 300~seconds and
reattempt the focus sequence. This step will be repeated until the
weather clears. If the cloudy weather lasts long enough, a new field
may be selected by \braind\ in the meantime.

Every morning, all of the FITS images from the previous night
(typically about 40~GB per telescope on a clear night) are archived to
a local RAID server in the main control building. Once every month or
so, the data are copied to a set of hard drives which are physically
brought to Academia Sinica. The data are copied to a local data
server, and then transfered to the Taiwan National Center for
High-Performance Computing\footnote{\url{http://www.nchc.org.tw/}} for
backup to tape. Finally, the data are copied over the internet to a
second RAID server located at National Central University. We then
have three copies of the data in separate locations for secure,
redundant data storage. The data are subsequently analyzed on using
two computer clusters (at Academia Sinica and National Central
University) with a custom photometry package that will be described in
\citet{kiwiphot}.

\subsection{The GRB Alert Response System}
\label{sec:grb}
The TAOS GRB response system consists of two components,
the \forwarder\ and the GRB daemon \grbd. The \forwarder\ simply
listens to a socket connected to the GCN, and passes the information
on to \grbd. The purpose of the \forwarder\ is to run on a machine with
a more reliable internet connection than that available at Lu-Lin
Observatory. If the internet connection to Lu-Lin is down, the
\forwarder\ still receives the GCN packet and keeps trying to forward
the packet to \grbd\ until it succeeds. We currently run
the \forwarder\ on a computer at the Institute for Astronomy and
Astrophysics at Academia Sinica in Taipei.

When \grbd\ receives a GCN alert packet from the \forwarder, it parses
the message and determines whether the reported event can be followed
up by the TAOS system. If the GRB event is observable,
\grbd\ calculates an observation schedule and passes the request on to
\braind, which will interrupt any zipper mode operations and schedule
the GRB follow-up observations.

TAOS has the advantage of multiple telescopes operating at the same
site, which makes it possible to use different exposure times for the
same event. Our observing scheme is optimized to have at least one
telescope looking for fine structure in the light curves of bright
events with short exposures, and two telescopes taking longer
exposures for sensitivity to fainter afterglows. In the future, we may
schedule one telescope to observe in zipper mode at 5~Hz to look for
very bright and rapidly varying transients. To date, 10~GRB alerts
were responded to by the TAOS system, and two afterglows
(R = 16.5 and 16.8) have been detected \citep{grb}.

\section{Field Selection}
\label{sec:fields}
In order to meet the survey requirements discussed in
\sect{sec:design}, a set of fields needed to be found which had
sufficient numbers of stars. This task was complicated by the fact
that the TAOS survey would be using zipper mode readout, as the
brighter stars will leave significant streaks in the images. It was
therefore desirable to minimize the numbers of bright stars in the
fields. A set of 167~fields was found, with varying characteristics
such as number of stars, ecliptic longitude and latitude. Information
on the TAOS fields can be found on the TAOS
website\footnote{\url{http://taos.asiaa.sinica.edu.tw/taosfield/}}.
Note that this set is much larger than is needed, however a large set
was generated for the first phase of the survey in order to evaluate
the fields in terms of signal-to-noise as a function of parameters
such as sky background, stellar density, etc. After analysis of the
first data set, a small subset of these fields will be selected for
further observations.

The fields were selected by using the USNO~B1.0 catalog
\citep{2003AJ....125..984M}. Initially, lists of stars were generated
in 300$\times$300 arc second tiles. These tiles covered all of the sky
with declinations in the range of $-22^\circ < \delta < 68^\circ$,
which corresponds to elevation angles higher than $45^\circ$ at the
Lu-Lin site.  After generating over 4.6~million tiles, these smaller
sections of the sky were merged into larger $1.7^\circ \times
1.7^\circ$ lists of stars. The center of each small tile (except for
those near the declination limits) became the center of the larger
field.  This process ensured overlap, allowing for the ability to
maneuver around the bright stars ($R < 9$) in the sky. Over four
million of the large fields were generated.

A subset of these fields was then selected according to the following
characteristics:
\begin{itemize}
\item Maximum number of stars in the field with $R < 15$.
\item Minimum number of bright stars ($R \leq 9$).
\item Maximum coverage of the ecliptic.
\item Some control fields away from the ecliptic.
\item No nebulae or other astronomical phenomena.
\end{itemize}

These requirements were used to narrow down the list of viable
fields. The field list was refined using a sorting routine to select
fields with varying star densities. The main selection requirement was
minimizing the streaks in the zipper mode images while not
significantly compromising on the total number of stars. Candidate
fields that did not meet these criteria were cut from the list.
Fields with nebulae or other phenomena were removed by checks of
Digital Sky Survey\footnote{\url{http://archive.stsci.edu/dss/}} (DSS)
images. The DSS images were also used to verify the accuracy of the
USNO~B1.0 catalog, particularly to ensure that there are no bright
stars in the field that do not appear in the catalog.

Note that it was impossible to meet the stellar density requirement in
all of the fields, as most of the fields along the ecliptic are very
sparse (some of these fields have fewer than 100~stars). However, care
was taken to find the densest fields available in these sparse
regions, and it should be noted that the sparsest fields are visible
mostly during the rainy summer months when observing is limited. Also
note that high priority is given to observations of the densest fields
in order to maximize the amount of useful data collected, and that the
sparser fields are only observed when necessary due to the season and
phase of the moon. The typical number of stars in the fields observed
in our first data set was thus 300 -- 500~stars \citep{kiwiapjl}.

\begin{figure}[bt]
\begin{center}
\plotone{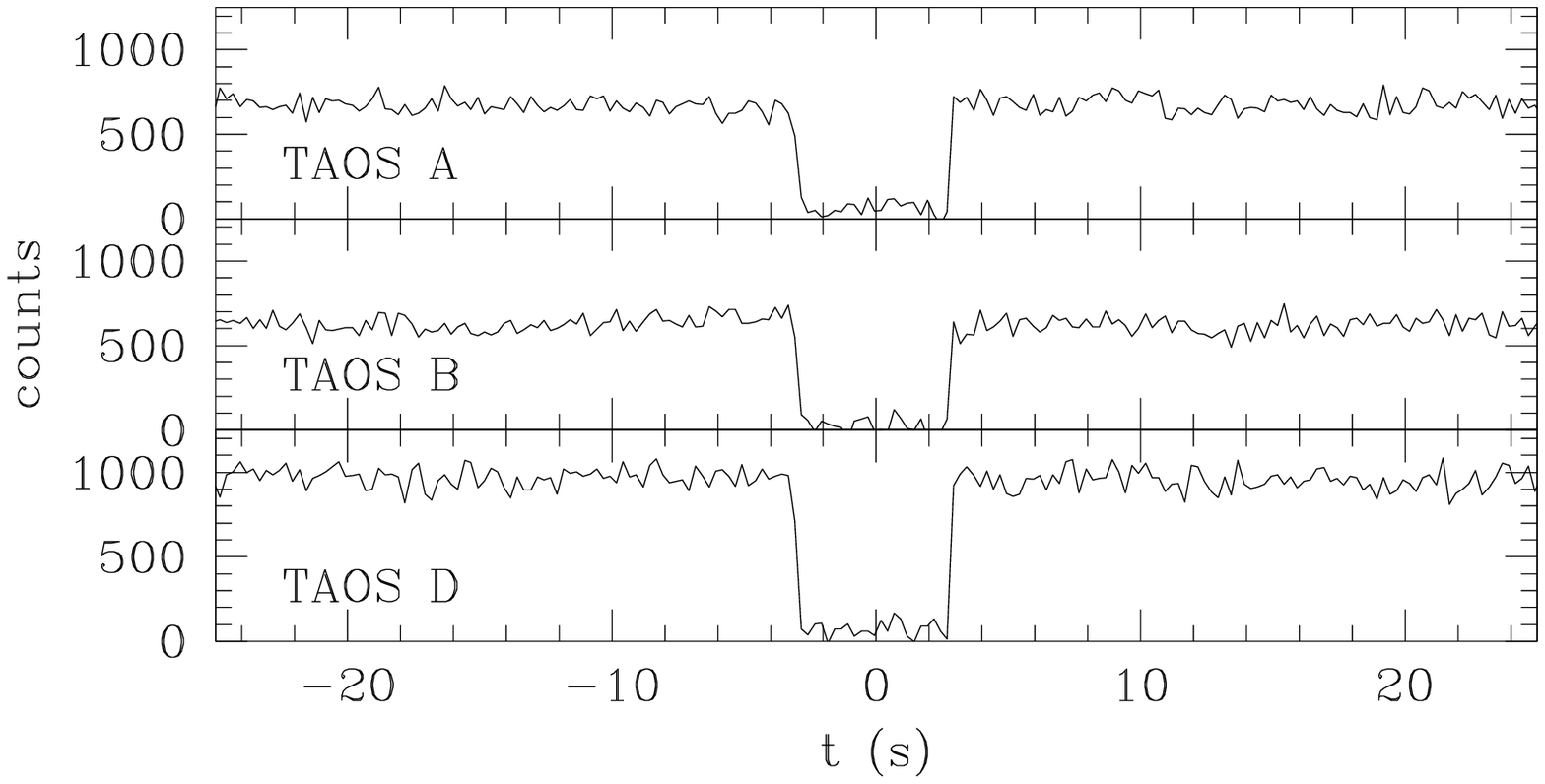}
\caption[]{Lightcurves from three TAOS telescopes of the occultation
  of the star TYC~076200961 ($V = 11.83$) by the asteroid Iclea ($V =
  14.0$, diameter~$= 97$~km) at UTC 12:35 February 6, 2006.}
\label{fig:iclea}
\vspace{.15in}
\plotone{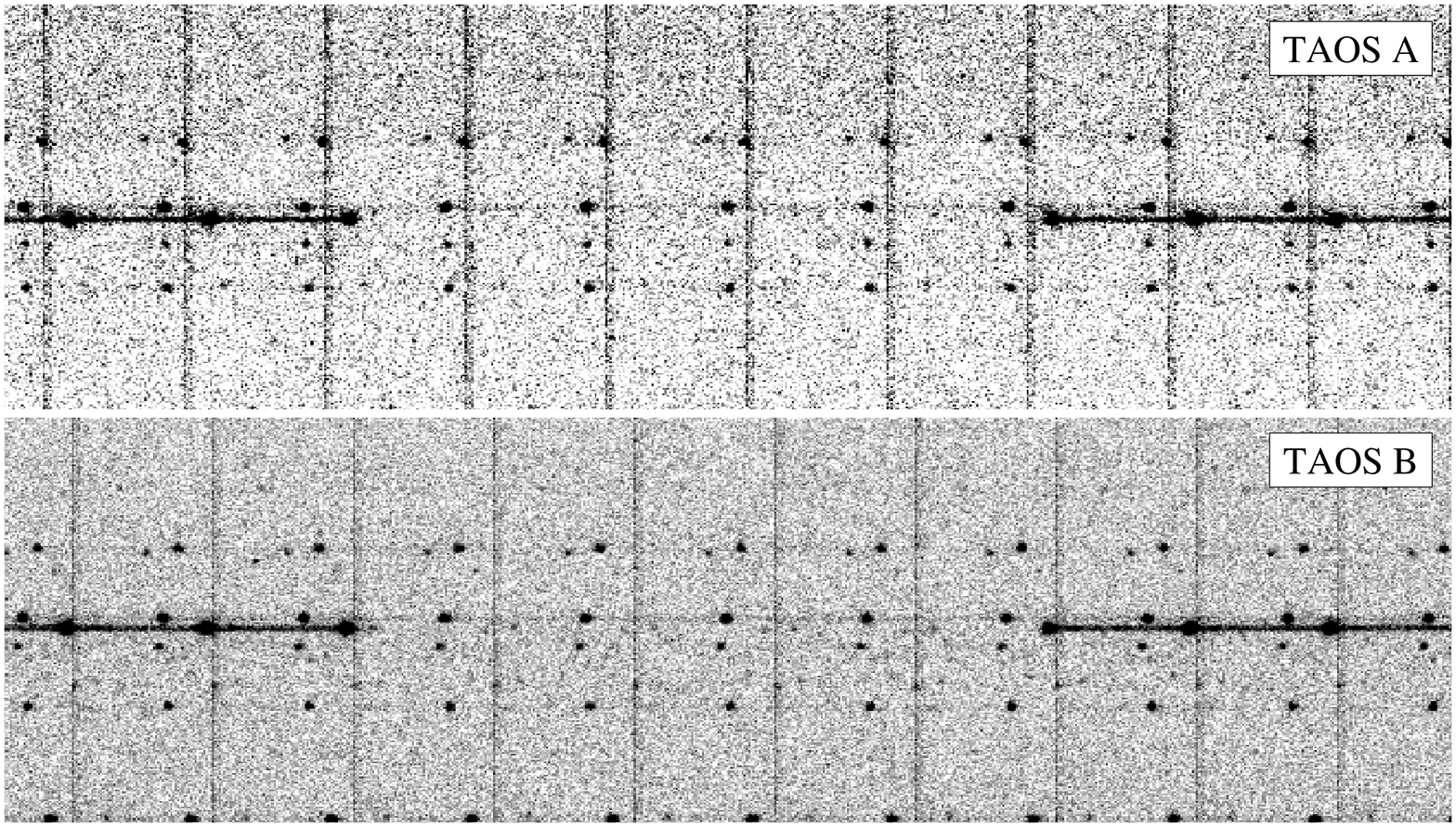}
\caption[]{Zipper mode sub-images from telescopes TAOS~A and TAOS~B of
  the occultation of the star HIP 050535 ($V = 8.46$) by the asteroid
  Klemola ($V = 15.7$, diameter~$= 31$~km) at UTC 12:10 June 5,
  2004. Each sub-image is an exposure of 0.25 seconds, with time
  increasing to the right. The vertical lines visible in each set of
  images are caused by a slight excess in dark current from the CCD
  substrate, and indicate the boundaries of the row blocks. These
  images were taken before \guided\ was implemented, so the two
  telescopes have different pointing offsets and the occulted star is
  in a different location in the row block.}
\label{fig:klemola}
\end{center}
\end{figure}

\section{System Performance}

To test the overall performance of the system we have observed several
predicted asteroid occultations\footnote{The asteroid occultation
  observations were based on the predictions of Dr. Isao Sato of the
  Nakano Star Gazers Club.}. \fig{fig:iclea} shows the lightcurves of
the occultation of the star TYC~076200961 ($V = 11.83$) by the
asteroid \emph{Iclea} ($V = 14.0$, diameter~$= 97$~km) at UTC 12:35
February 6, 2006. The event was observed by operating the three
functioning telescopes at 4~Hz, and the event, which lasted
approximately 6~seconds, is clearly visible, indicating that the
system is indeed capable of accurate, high-speed
imaging. \fig{fig:klemola} shows a series of zipper mode images of the
occultation of the star HIP~050535 ($V = 8.46$) by the asteroid
\emph{Klemola} ($V = 15.7$, diameter~$= 31$~km) at UTC 12:10 June 5,
2004. This was a short duration event, lasting only 1.3~seconds.

These asteroid occultation events clearly illustrate that the TAOS
system is capable of detecting short duration occultation events with
multiple telescopes. The system has been collecting data for over
three years with three telescopes, and the fourth telescope came
online in August 2008. Since February of 2005, we estimate TAOS has
made over $5 \times 10^{10}$~individual photometric measurements, and
analysis of the first two years of data has been completed
\citep{kiwiapjl}.

\acknowledgements This work was supported in part by the National
Science Foundation under grant AST-0501681 and by NASA under grant
NNG04G113G, both at the Harvard College Observatory. Work at Academia
Sinica was supported in part by the thematic research program
AS-88-TP-A02. The work at National Central University was supported by
the grant NSC 96-2112-M-008-024-MY3. SLM's work was performed under
the auspices of the U.S. Department of Energy by Lawrence Livermore
National Laboratory in part under Contract W-7405-Eng-48 and by
Stanford Linear Accelerator Center under Contract
DE-AC02-76SF00515. Work at Yonsei was supported by the KRCF grant to
Korea Astronomy and Space Science Institute. KHC's work was performed
under the auspices of the U.S. Department of Energy by Lawrence
Livermore National Laboratory in part under Contract W-7405-Eng-48 and
in part under Contract DE-AC52-07NA27344. MJL wishes to thank Jeff
Klein at the University of Pennsylvania for valuable advice regarding
the system hardware development. MES wishes to thank Jeff Goldader and
Gary Bernstein at the University of Pennsylvania for advice on the
TAOS field selection.

\end{document}